\documentclass[aps,pra,eqsecnum,twocolumn,showpacs,superscriptaddress]{revtex4}

\usepackage{amsfonts,amssymb,amsmath}
\usepackage{mathtools}
\usepackage[]{graphics,graphicx,epsfig}
\usepackage{amsthm}

\begin{document}

\title{Coherent states of composite bosons}

\author{Su-Yong \surname{Lee}}
\email{cqtlsy@nus.edu.sg}
\affiliation{Centre for Quantum Technologies, National University of Singapore, 3 Science Drive 2, 117543 Singapore, Singapore}

\author{Jayne \surname{Thompson}}
\affiliation{Centre for Quantum Technologies, National University of Singapore, 3 Science Drive 2, 117543 Singapore, Singapore}

\author{Pawe{\l} \surname{Kurzy\'nski}}
\affiliation{Centre for Quantum Technologies, National University of Singapore, 3 Science Drive 2, 117543 Singapore, Singapore}
\affiliation{Faculty of Physics, Adam Mickiewicz University, Umultowska 85, 61-614 Pozna\'{n}, Poland}

\author{Akihito \surname{Soeda}}
\affiliation{Department of Physics, Graduate School of Science, University of Tokyo, 7-3-1, Hongo, Bunkyo-ku, Tokyo, Japan}
\affiliation{Centre for Quantum Technologies, National University of Singapore, 3 Science Drive 2, 117543 Singapore, Singapore}

\author{Dagomir \surname{Kaszlikowski}}
\email{phykd@nus.edu.sg}
\affiliation{Centre for Quantum Technologies, National University of Singapore, 3 Science Drive 2, 117543 Singapore, Singapore}
\affiliation{Department of Physics, National University of Singapore, 2 Science Drive 3, 117542 Singapore, Singapore}

\date{\today}

\begin{abstract}
We present a systematic analysis on coherent states of composite bosons consisting of two distinguishable particles. 
By defining an effective composite boson (coboson) annihilation operator, we derive its eigenstate and commutator.
Depending on the elementary particles comprising the composite particles, we gauge the resemblance between this eigenstate and traditional coherent states through typical measures of nonclassicality, such as quadrature variances and Mandel's Q parameter. 
Furthermore, we show that the eigenstate of the coboson annihilation operator is useful in estimating the maximum eigenvalue of the coboson number operator.
\end{abstract}

\maketitle

\section{Introduction}

Composite particles are foundationally important to physics at all scales ranging from Cooper pair formation in superconductors, through Bose-Einstein condensation, and beyond.
Indeed at the cosmological scale we relabel all ordinary matter as ``baryonic matter" in deference to the fact that protons and neutrons dominate the standard model contribution to the energy budget of the universe. To characterize the phenomenology of these composite particles we often wish to benchmark how closely they imitate elementary bosons (or fermions). If a composite particle exhibits ideal bosonic behavior,  then it has an integer total spin, and as a corollary of the spin statistics theorem the wave function of a multiparticle composite boson (coboson) state is symmetric under the exchange of any two cobosons. 
Prominent examples include hydrogen atoms, which consist of a proton and an electron;  semiconductor excitons, which are composed of electron-hole pairs;
and biphotons generated by spontaneous parametric down-conversion, which can exhibit composite behavior even when the single photon subsystems are spatially separated and noninteracting \cite{L05}. 
The latter have engendered speculation about the origin of composite behavior. 

Formation of composite particles has recently attracted a lot of attention from entanglement and many-body physics communities
 \cite{CBD08, KG08, BRDLBG08, CDD09, COW10, RKCSK11, C11, CSC11, TMB11, TBM12, KRSCK12, Tichy12, GM12, GM13, T13, CK13, TBM13, TBM131}.
Quantum entanglement between the two constituent fermions as quantified by the purity of the reduced single particle subsystems is believed to be the resource behind ideal bosonic behavior of composite particles \cite{L05}. Analysis of composite systems shows that extremizing the purity improves the bosonic behavior \cite{COW10,TBM12, RKCSK11}. The degree to which composite bosons imitate elementary bosons can also be analytically quantified by studying the effect of adding and subsequently subtracting a single coboson \cite{KRSCK12} and by novel interference schemes for many particle coboson states \cite{Tichy12}. 

In the context of innovative research on many particle coboson states and interferometry we are motivated to ask whether composite bosons (cobosons) can form coherent states. Coherent states hold special significance in interferometry experiments. In quantum optics coherent states are pure states which are separable after passing through a beam splitter. A dual motivation for this research is that coherent states are an important benchmark for studying nonclassicality in quantum optics. From this perspective formulating coherent states of cobosons will give us new tools for studying coboson statistics.  For instance, a coherent state of a radiation field exhibits Poissonian photon counting statistics as well as minimum uncertainty in position-momentum phase space \cite{G63}, whereas Fock states demonstrate sub-Poissonian photon counting statistics which are a hallmark of nonclassicality. In this paper we use the framework of effective coboson operators to formulate coherent states and study their counting statistics and quadrature variances. 
We then examine a potential application for coherent states of cobosons within the proposed framework of maximum occupation number (MON),
where the MON is a restriction on the amount of composite particles that can condense \cite{RVPP02}.

This paper is organized as follows. Section II begins with our definition of the effective composite boson (coboson) annihilation operator and provides the eigenstate and the commutator.
In Sec. III, we investigate the characteristics of the effective coboson operator and the eigenstate in quadrature variances and Mandel's Q parameter. Then we consider the usefulness of the eigenstate for estimating the maximum occupancy number in Sec. IV. 

\section{Effective composite boson annihilation operator : Eigenstate, and Commutator} 
A composite boson is composed of two distinguishable particles A and B, in which both particles are either bosons or fermions.
The wave function is decomposed into Schmidt modes as follows \cite{L05},
\begin{eqnarray}
\Psi(x_A, x_B)=\sum^{\infty}_{p=0}\sqrt{\lambda_p}\phi^{(A)}_p(x_A)\phi^{(B)}_p(x_B).
\end{eqnarray}
In the second quantization representation, the corresponding number state is generated by a creation operator, $\hat{c}^{\dag}=\sum^{\infty}_{p=0}\sqrt{\lambda_p}\hat{a}^{\dag}_p\hat{b}^{\dag}_p$,
acting on the vacuum.
The number state is defined as 
\begin{eqnarray}
|n\rangle \equiv \frac{1}{\sqrt{\chi_n}}\frac{\hat{c}^{\dag n}}{\sqrt{n!}}|0\rangle,
\end{eqnarray}
where $\chi_n$ is a normalization constant \cite{L05}. The corresponding annihilation operator acts on $|n\rangle$ as
\begin{eqnarray}
\hat{c}|n\rangle=\sqrt{\frac{\chi_n}{\chi_{n-1}}}\sqrt{n}|n-1\rangle +|\epsilon_n\rangle,
\end{eqnarray}
where $\langle n-1|\epsilon_n\rangle=0$, and $n>1$.
The correction term $|\epsilon_n\rangle$ has the property
\begin{eqnarray}
\langle \epsilon_n|\epsilon_n\rangle=1-n\frac{\chi_{n}}{\chi_{n-1}}+(n-1)\frac{\chi_{n+1}}{\chi_n},
\end{eqnarray}
where $\chi_{n\pm 1}/\chi_n\rightarrow 1$ for perfect bosonic property \cite{L05}. According to the comprising particles, i.e. elementary bosons (fermions), 
we have the relation $\chi^B_{n+1}/\chi^B_n\geq 1$ ($\chi^F_{n+1}/\chi^F_n\leq 1$) \cite{L05}.
With the assumption of small $\langle \epsilon_n|\epsilon_n\rangle$ , it follows that 
\begin{equation}
\hat{c} \equiv \sum^{\infty}_{n=0} |n\rangle \langle n+1| f_{n+1}+O(\epsilon_n),
\end{equation}
where state $|n\rangle$ represents the number of composite bosons, and $f_{n+1}=\sqrt{\frac{\chi_{n+1}}{\chi_n}}\sqrt{n+1}$,
which goes to $\sqrt{n+1}$ when the coefficients $\lambda_p$ in Eq. (2.1) are equalized for the infinite number of mode pairs.

We assume that the corresponding eigenstate of the annihilation operator, Eq. (2.5), which is conventionally known as a coherent state, can be expanded in terms of the complete basis $\{ |n\rangle : n\in {\Bbb N}\}$,
\begin{equation}
|\psi\rangle = \frac{1}{\sqrt{\sum^{\infty}_{n=0}|\psi_n|^2}}\sum^{\infty}_{n=0} \psi_n |n\rangle,
\end{equation}
with the condition
\begin{eqnarray}
\hat{c}|\psi\rangle &=& \frac{1}{\sqrt{N}}\sum^{\infty}_{n=0} \psi_n \hat{c}|n\rangle = \frac{1}{\sqrt{N}}\sum^{\infty}_{n=1} \psi_n f_n |n-1\rangle \notag\\
&=&\frac{\gamma }{\sqrt{N}} \sum^{\infty}_{n=0} \psi_n |n\rangle=\gamma|\psi\rangle,
\end{eqnarray}
where $N=\sum^{\infty}_{n=0}|\psi_n|^2$.
This implies that $\gamma \psi_{n-1} = f_n \psi_{n}$ and therefore
\begin{equation}
\psi_n = \frac{\gamma^n}{\prod^{n}_{i = 1} f_i}\psi_0.
\end{equation}
At this stage we take $f_0 = 1$. 
For clarity we include the explicit formulation of the coherent state of cobosons:
\begin{equation}
|\psi\rangle = \frac{1}{\sqrt{\cal N}} \sum^{\infty}_{n=0} \frac{\gamma^n}{\prod^{n}_{i = 0 } f_i} |n\rangle,
\label{eq:coherentstate}
\end{equation}
where the normalization constant $\cal N$ is 
\begin{equation}
{\cal N} = \frac{N}{|\psi_0|^2}= \sum_n |\psi_n/\psi_0|^2=\sum^{\infty}_{n=0} \frac{|{\gamma}|^{2n}}{\prod^{n}_{i = 0} |{f_i}|^2} .
\label{eq:normconstant}\end{equation}
An implicit caveat on this construction is that the normalization constant ${\cal N}$ must be finite. Series (\ref{eq:normconstant}) converges when the absolute value of successive terms in the sequence $\{|\psi_n|: n\in {\Bbb N}\}$, is monotonically decreasing in the limit $n \rightarrow \infty$ according to
\begin{equation}
\lim_{n\rightarrow \infty} |\psi_n/\psi_{n-1}| \le 1,
\end{equation}
which implies $|\gamma| \le \lim_{n\rightarrow \infty} |f_n|$.
This indicates that there is a collection of well defined eigenstates for any anihilation operator satisfying $\lim_{n\rightarrow \infty} |f_n| > 0$ .

Based on the property of the effective coboson operator, we can derive the commutator,
\begin{eqnarray}
\left[\hat{c}, \hat{c}^{\dagger}\right] &= &  \sum^{\infty}_{n=0} |f_{n+1}|^2|n\rangle \langle n| \notag -\sum^{\infty}_{n=0}|f_{n+1}|^2|n+1\rangle \langle n+1| \notag\\
&= & |f_1|^2 |0\rangle \langle0|- \sum^{\infty}_{n=1} \left( |f_{n}|^2 - |f_{n+1}|^2 \right) |n\rangle \langle n|,\notag\\
\end{eqnarray}
where the commutator reduces to the identity in the familiar case $f_n=\sqrt{n}$.
For the eigenstate (\ref{eq:coherentstate}), the commutator is evaluated as
\begin{eqnarray}
  \langle \left[\hat{c}, \hat{c}^{\dagger}\right] \rangle 
&=& \frac{1}{\cal N} \left[ |f_1|^2-\sum^{\infty}_{n=1} \frac{|\gamma|^{2n}}{ \prod^{n}_{i=1} |f_i|^{2}} \left( |f_{n}|^2 - |f_{n+1}|^2 \right)  \right].\notag \\
\end{eqnarray}
The commutation relation is also given by $\left[\hat{c}, \hat{c}^{\dagger}\right] =1+s\Delta$
for cobosons comprised of elementary bosons ($s=1$) or fermions ($s=-1$), where $\Delta=\sum\lambda_p(\hat{a}^{\dag}_p\hat{a}_p+\hat{b}^{\dag}_p\hat{b}_p)$
 ($\hat{a}_p$ and $\hat{b}_p$ are elementary boson or fermion annihilation operators). 
 Thus, the expectation value of the commutator, $\langle \left[\hat{c}, \hat{c}^{\dagger}\right] \rangle$, will be $>1~(<1)$ when the cobosons are comprised of elementary bosons (fermions) \cite{L05}.
As an example, we can consider the classical annihilation operator \cite{LNK13}, i.e., the scenario where the composite particles are distinguishable, 
which is prescribed by setting $f_n=1$ for all $n$.
Then, the expectation value of the commutator is equal to $ \langle \left[\hat{c}, \hat{c}^{\dagger}\right] \rangle=1-|\gamma|^2 \leq 1$ 
due to the relation $|\gamma| \le \lim_{n\rightarrow \infty} |f_n|=1$.

\section{Characteristics of the effective coboson operator and corresponding coherent state}
In this section we apply well known measures of nonclassicality to the coboson annihilation operator and corresponding coherent state; including quadrature variances and Mandel's Q parameter \cite{M79}.
We derive effective quadrature variances and Mandel's Q parameter with an expectation value of the commutator, $\langle\left[\hat{c}, \hat{c}^{\dagger}\right]\rangle $.
Note that we cannot say anything about nonclassical properties of the composite bosons.

\begin{figure}
\centerline{\scalebox{0.31}{\includegraphics[angle=270]{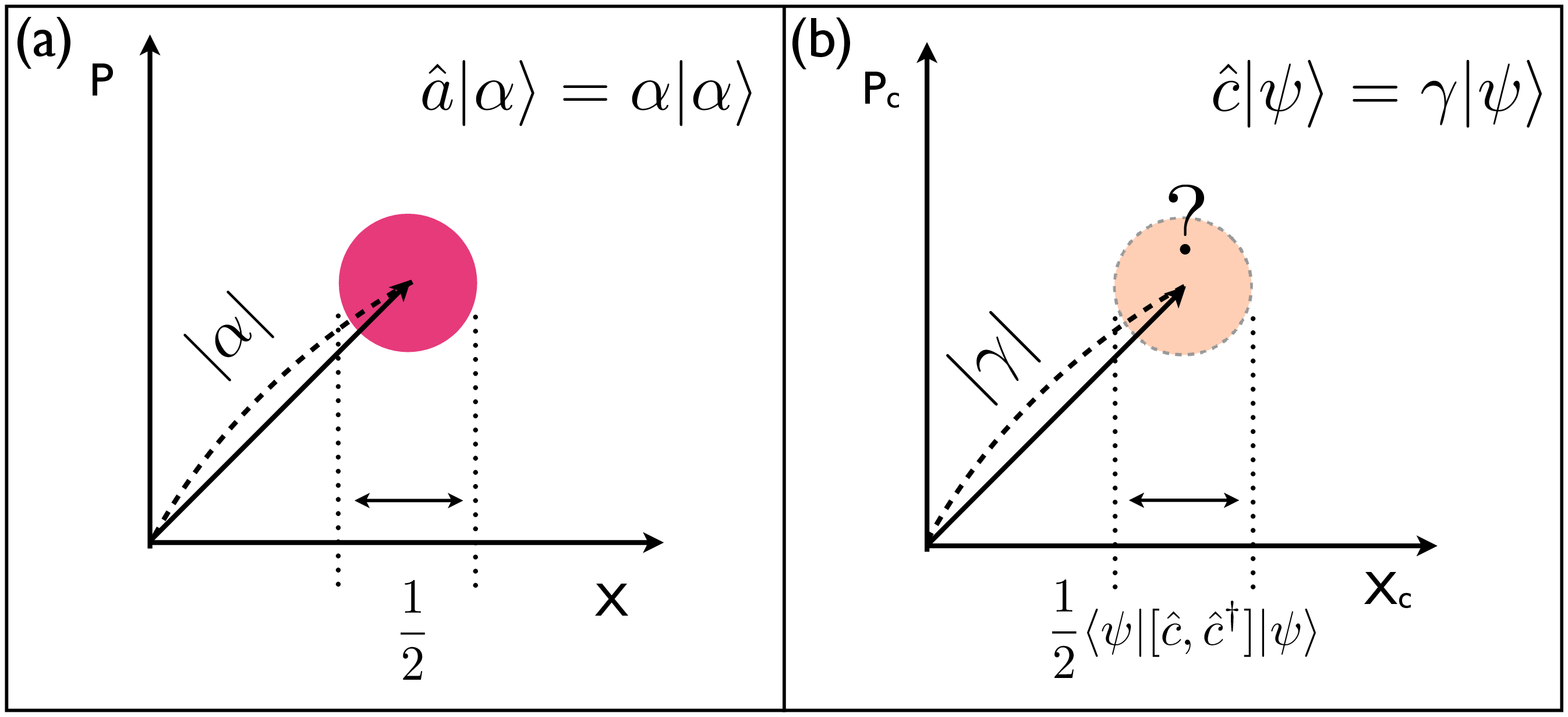}}}
\vspace{-1.1in}
\caption{Position-momentum phase-space representations of coherent states for (a) elementary bosons (e.g., photons) and (b) composite bosons.
}
\label{fig:fig1}
\end{figure}

\subsection{Quadrature variances}
The position-momentum phase-space representation of this coherent state of cobosons is depicted in Fig. 1, which facilitates a direct comparison with photonic coherent states.
For photonic states, quadrature variances in position-momentum phase space are defined as
\begin{eqnarray}
(\Delta X)^2=\langle \hat{X}^2\rangle -\langle \hat{X}\rangle^2,~(\Delta P)^2=\langle \hat{P}^2\rangle -\langle \hat{P}\rangle^2,
\end{eqnarray}
where $\hat{X}=\frac{\hat{a}+\hat{a}^{\dag}}{2}$ and $\hat{P}=\frac{\hat{a}-\hat{a}^{\dag}}{2i}$.
In this scenario, the quadrature variances are $(\Delta X)^2=(\Delta P)^2=1/4$,
such that coherent states have a quadrature minimum uncertainty of $(\Delta X)^2(\Delta P)^2=1/16$.
Any state which is characterized with a quadrature variance less than $1/4$, i.e., $(\Delta X)^2 <1/4$ or $(\Delta P)^2<1/4$, is called squeezed.
We consider the variances for the effective coboson annihilation operator $\hat{c}$, $(\Delta P_{e\!f\!f})^2=(\Delta X_{e\!f\!f})^2$, and

\begin{eqnarray}
(\Delta X_{e \!f \!f})^2
&=&\frac{1}{4}\langle \hat{c}^2+ \hat{c}^{\dag 2}+ \hat{c} \hat{c}^{\dag}+\hat{c}^{\dag}\hat{c}\rangle
-\frac{1}{4}\langle \hat{c}+\hat{c}^{\dag}\rangle^2 \notag\\
&=&\frac{1}{4}(\gamma^2+\gamma^{\ast 2}+2|\gamma|^2+\langle [\hat{c} ,\hat{c}^{\dag}]\rangle )-\frac{1}{4}(\gamma+\gamma^{\ast})^2 \notag\\
&=&\frac{1}{4}\langle [\hat{c} ,\hat{c}^{\dag}]\rangle,
\end{eqnarray}
where $\hat{X}_{e \!f \!f}=(\hat{c}+\hat{c}^{\dag})/2$.
It follows that the eigenstate of the effective coboson operator can have a quadrature variance of $1/4$ at $\langle [\hat{c} ,\hat{c}^{\dag}]\rangle=1$. 
The coboson states comprised of elementary bosons (fermions) can attain a higher (lower) quadrature variance than $1/4$.

\subsection{Mandel's Q parameter}
Photon counting statistics is a simple measure for distinguishing nonclassical states from classical states. 
The photon counting statistics which is called Mandel's Q parameter, is defined as
\begin{equation}
Q=\frac{(\Delta \hat{n})^2-\langle \hat{n}\rangle}{\langle \hat{n}\rangle}=\frac{  \langle \hat{n}^2\rangle-
\langle \hat{n}\rangle^2-\langle \hat{n}\rangle}{\langle \hat{n}\rangle},
\end{equation}
where $\hat{n}=\hat{a}^{\dagger}\hat{a}$. Quantum states belonging to the range $-1\leq Q <0$ have a sub-Poissonian distribution, 
whereas states belonging to the range $Q\geq0$ have a Poissonian or super-Poissonian distribution. 
Coherent states are an important benchmark for nonclassicality; they are conventionally characterized by $Q = 0$ and manifest a Poissonian distribution.
Here we can only state the distribution of statistics for coherent states of composite bosons, depending on the comprising particles, i.e., bosons or fermions.
Using the properties of the effective coboson annihilation operator $\hat{c}$ and the eigenstate $|\psi\rangle$, i.e.,
$\langle \hat{c}^{\dagger}\hat{c}\rangle = |\gamma|^2$ and  $\hat{c} \hat{c}^{\dagger} =\hat{c}^{\dagger} \hat{c}+ [\hat{c}, \hat{c}^{\dagger}]$, we can derive an effective Mandel's Q parameter,
\begin{eqnarray}
Q _{e\!f\!f}&=& \frac{\langle \hat{c}^{\dagger} \hat{c} \hat{c}^{\dagger} \hat{c}\rangle
- \langle \hat{c}^{\dagger} \hat{c}\rangle^2 }{\langle \hat{c}^{\dagger} \hat{c}\rangle}-1 \notag\\
&=& \frac{|\gamma|^4+ |\gamma|^2 \langle [\hat{c} ,\hat{c}^{\dag}]\rangle  - |\gamma|^4 }{|\gamma|^2}-1\notag\\
&=& \langle [\hat{c} ,\hat{c}^{\dag}]\rangle-1,
\end{eqnarray}
where the value of $Q_{e\!f\!f}$ can be equal to zero at $\langle [\hat{c} ,\hat{c}^{\dag}]\rangle=1$.
The coboson states of elementary bosons (fermions) can attain positive (negative) values which correspond to super-Poissonian (sub-Poissonian) statistics.

\section{Application: Estimation of the maximum occupancy number}
In this section we consider an application. 
We demonstrate how coherent states of cobosons can be used to estimate the maximum eigenvalue of the coboson number operator $\hat n=\hat c^{\dagger} \hat c$. Physically this eigenvalue corresponds to the maximum number of cobosons that can occupy a single mode. This number limits potential realizations of Bose-Einstein condensates (BECs) of composite particles, since it gives a natural restriction on the amount of composite particles that can condense.
The notion of composite bosons applies to situations in which one assumes that there exists an effective Hamiltonian for which bounded states of two (or more) fermions form an eigenbasis.


For elementary bosons the maximum occupancy number is infinite, since the corresponding number operator is not bounded from above, and hence there is no restriction on the number of particles that can condense. However, in the case of composite particles there might be such a restriction. This problem was studied in Ref. \cite{RVPP02} using the approach of the ground-state energy of a fermionic generalized pairing problem. Here, we show that a lower bound on the maximum occupancy number naturally stems from the properties for coherent states of cobosons.

We first note that for any Hermitian operator $\hat{A}$, one has
\begin{equation}
\langle\phi|\hat{A}|\phi\rangle \leq a_{max},
\end{equation}
for any state $|\phi\rangle$, where $a_{max}$ is the maximal eigenvalue of $\hat{A}$. Next, note that for the coherent state $|\psi\rangle$ (\ref{eq:coherentstate}) one has
\begin{equation}
\hat{c}|\psi\rangle=\gamma |\psi\rangle,
\end{equation}
and hence
\begin{equation}\label{bound}
\langle \psi| \hat n |\psi\rangle = |\gamma|^2.
\end{equation}
Therefore, we have $|\gamma|^2 \leq n_{max}$.

Furthermore, under the assumption that states of cobosons are eigenstates of the effective Hamiltonian, coherent states of cobosons have to span the effective eigenbasis describing the physical system.
As a consequence of this assumption, the bound, (\ref{bound}), is true for every state. In addition, we showed that $|\gamma| \le \lim_{n\rightarrow \infty} |f_n|$, therefore we conclude that\begin{equation}
\lim_{n\rightarrow \infty} |f_n|^2 \approx n_{max}.
\end{equation}
For $f_n=1$ for all $n$, i.e., distinguishable particles, the lower bound is obtained as $ |\gamma|^2\leq n_{max}=1$.

\section{Conclusion}
We have conducted a systematic analysis of coherent states of composite bosons. 
For these states we have described the corresponding effective quadrature variances and Mandel's Q parameter with the expectation value of the commutator, $\langle[\hat{c},\hat{c}^{\dag}]\rangle$. 
According to whether the coherent states of cobosons are comprised of elementary bosons or fermions, i.e., $\langle[\hat{c},\hat{c}^{\dag}]\rangle \geq 1$ or $\leq 1$,
the effective quadrature variances can achieve higher or lower quadrature variances than those
for coherent states of elementary bosons, and the effective Mandel's Q parameter can have super or sub-Poissonian statistics, whereas coherent states of elementary bosons have
Poissonian statistics.
As an application, we have derived the lower bound on the MON from coherent states of cobosons.
As a further work, we can examine Gaussian states which include not only squeezed coherent and displaced (squeezed) thermal states
but also the superposition of a thermal field on a Gaussian one \cite{MMS02}.


\begin{acknowledgments}
We would like to thank the anonymous referee for providing us with constructive comments.
This work was supported by the National Research Foundation and Ministry of Education in Singapore.
\end{acknowledgments}


\end{document}